\documentclass[twocolumn,showpacs,preprintnumbers,aps]{revtex4}
\usepackage[dvips]{graphicx}
\usepackage{array}
\usepackage{calc}

\newcommand{\PBS}[1]{\let\temp=\\#1\let\\=\temp}

\begin{document}
\preprint{PHYSICAL REVIEW A, {\bf 66}, 022115 (2002)}
\title{Relativistic coherent states and charge structure of the
coordinate and momentum operators}

\author{B.I. Lev} \altaffiliation[Also at ]{Physics
Department, Taras Shevchenko Kiev University}
\email{lev@iop.kiev.ua}
\author{A.A. Semenov}%
 \email{sem@iop.kiev.ua}
\affiliation{Institute of Physics, National Academy of Sciences of
Ukraine, 46 Nauky pr, Kiev 03028, Ukraine}%
\author{C.V. Usenko}
\email{usenko@ups.kiev.ua} \affiliation{Physics Department, Taras
Shevchenko Kiev University, 6 Academician Glushkov pr, Kiev 03127,
Ukraine}
\author{J.R. Klauder} \email{klauder@phys.ufl.edu}
\affiliation{Departments of Physics and Mathematics, University of
Florida, Gainesville, Florida 32611}
\date{19 April 2002}
\begin{abstract}
We consider relativistic coherent states for a spin-$0$ charged
particle that satisfy the next additional requirements: (i) the
expected values of the standard coordinate and momentum operators
are uniquely related to the real and imaginary parts of the
coherent state parameter $\alpha$; (ii) these states contain only
one charge component. Three cases are considered: free particle,
relativistic rotator, and particle in a constant homogeneous
magnetic field. For the rotational motion of the two latter cases,
such a description leads to the appearance of the so-called
nonlinear coherent states.
\end{abstract}
 \maketitle

\section{Introduction}
\label{s1}

Coherent states are currently very interesting in modern quantum
physics and in quantum technologies. They have a long history and
a very wide field of application. It was Schr\"odinger who first
considered these states for the harmonic oscillator in the early
years of quantum mechanics \cite{b1}. These states minimize the
uncertainty relations and one of their important properties is
temporal stability. It means that time evolution can be described,
in this case, by changing the parameter of the coherent states
\begin{equation}
\alpha\longmapsto \alpha\exp\left(-i\omega t\right). \label{f1}
\end{equation}

These states are eigenstates of the annihilation operator
\begin{equation}
\hat{a}=\frac{1}{\sqrt{2}}\left(\frac{\hat{q}}{\sigma}+
i\frac{\sigma}{\hbar}\hat{p}\right),\label{f2}
\end{equation}
where $\sigma=\sqrt{\hbar/m\omega}$ is a characteristic oscillator
length, with eigenvalues $\alpha$,
\begin{equation}
\hat{a}\left|\alpha\right\rangle=\alpha\left|\alpha\right\rangle.
\label{f3}
\end{equation}

Another significant property of the coherent states is the
resolution of unity \cite{b2, b3}. It means the existence of an
integral measure $d\mu(\alpha)$ such that the following condition
is satisfied:
\begin{equation}
\int \left|\alpha\right\rangle d\mu(\alpha) \left\langle
\alpha\right|=1. \label{f4}
\end{equation}
For the standard coherent states determined by Eq. (\ref{f3}) this
measure has a simple form:
\begin{equation}
d\mu(\alpha)=\frac{d\alpha^{\prime}d\alpha^{\prime\prime}}{\pi},
\label{f5}
\end{equation}
where $\alpha^{\prime}$ and $\alpha^{\prime\prime}$ are the real
and imaginary parts of the parameter $\alpha$.

These coherent states (we denote them as the standard coherent
states here) can be used not only for a harmonic oscillator but
for a wide enough class of the quantum systems. However, the
property of the temporal stability is intrinsic to the systems
which are unitarily equivalent to the harmonic oscillator. This
property can be conserved by a redefinition of the coherent states
\cite{b4, b5}.

A lot of different generalizations of the coherent states are well
known today. We note only one of them that will appear in our
consideration from an unexpected side. In \cite{b6} the authors
have shown that trapped electron can be described by means of
coherent states which are eigenstates of the deformed annihilation
operator $\hat{a}_f$ expressed in terms of the usual annihilation
and creation operators in the following way:
\begin{equation}
\hat{a}_f=\hat{a}f\left(\hat{a}^{\dag}\hat{a}\right), \label{f6}
\end{equation}
where $f$ is a certain function. Mathematical properties and a
physical sense of these states have been considered in detail in
\cite{b7}. The authors call these states nonlinear (or $f$)
coherent states. Analogous states that satisfy the condition of
temporal stability have been presented in \cite{b4, b5}.

The description of a relativistic quantum system by means of the
coherent states meets certain difficulties. First of all, it
should be noticed that relativistic quantum theory can be
developed in a consistent way with the second quantization method
only. It is possible to consider the one-particle sector of the
theory under conditions when particle pair creation is impossible
(free particle and particle in a constant magnetic field).
However, even in this case the coordinate (and, generally
speaking, the momentum) operator is not well defined \cite{b8, b9,
b10}. This reveals itself in the fact that eigenstates of this
operator contain states with different signs of charge. Therefore,
the coherent states defined by Eq. (\ref{f3}) contain eigenstates
of the Hamiltonian with different signs of charge as well. The
existence of such kind of states is prohibited by the charge
superselection rule \cite{b11}.

Coherent states for relativistic spin-$0$ and spin-$1/2$ particles
in a constant homogeneous magnetic field have been introduced in
\cite{b12}. These states are expanded in eigenstates of the
Hamiltonian with one sign of charge only. However, these states
are not eigenstates of the annihilation operator, and furthermore,
expected values of coordinate and momentum for these states are
not $\alpha^{\prime}\sqrt{2}\sigma $ and
$\alpha^{\prime\prime}\sqrt{2}\hbar /\sigma$, respectively.

The coherent states for charged spin-$0$ particles which we
consider in this work in detail, have been presented in
\cite{b13}. These states satisfy two additional conditions:
\begin{enumerate}
\item \label{i1} The coherent state parameter $\alpha$ is expressed
in terms of the mean value of the standard (not Newton -- Wigner)
coordinate and momentum in the usual way,
\begin{equation}
\alpha=\frac{1}{\sqrt{2}}\left(\frac{\bar{q}}{\sigma}+
i\frac{\sigma}{\hbar}\bar{p}\right).\label{f7}
\end{equation}
\item \label{i2} These states are expanded in eigenstates
of the Hamiltonian with one sign of charge only.
\end{enumerate}

The coherent states, which can be constructed by using Eq.
(\ref{f3}), satisfy condition (\ref{i1}) but do not satisfy
condition (\ref{i2}); the coherent states presented in \cite{b12},
in contrast, satisfy condition (\ref{i2}) but do not satisfy
condition (\ref{i1}).

In this work we consider a spin-$0$ charged particle in a constant
homogeneous magnetic field and two more simple examples: free
particle and relativistic rotator. These systems can be described
in the one-particle sector of the theory because particle pair
creation is absent there. However, the nontrivial charge structure
of the coordinate and momentum operators leads to the appearance
of some peculiarities in the evident form of the relativistic
coherent states and in the expected values of some observables.

\section{Statement of the problem and charge structure of operators}
\label{s2}

Consider a scalar charged particle in a constant homogeneous
magnetic field. Such a particle is described by the Klein--Gordon
equation:
\begin{equation}
-\hbar^2\partial_t^2\psi=\left[c^2\left(\mathbf{\hat{p}}-e\mathbf{A}
\left(\mathbf{\hat{r}}\right)\right)^2+m^2c^4\right]\psi.
\label{f7a}
\end{equation}
This is a second-order equation in time. Hence, in the present
form, one cannot interpret it as a Schr\"odinger equation. This
problem can be resolved, if by means of changing

\begin{eqnarray}
\psi=\frac{1}{\sqrt{2}}\left(\varphi+\chi\right),\label{f7b}\\
i\hbar\partial_t\psi=\frac{mc^2}{\sqrt{2}}\left(\varphi-\chi\right)\label{f7c}
\end{eqnarray}
one comes to the two-component wave function
\begin{equation}
\Psi=\left(\begin{array}{c}\varphi\\
\chi\end{array}\right).\label{f7d}
\end{equation}
Now, Eq. (\ref{f7a}) can be written as the Schr\"odinger equation
\begin{equation}
i\hbar\partial_t\Psi=\hat{H}^{mf}\Psi,\label{f7e}
\end{equation}
where the Hamiltonian $\hat{H}^{mf}$ has the following form:
\begin{equation}
\hat{H}^{mf}=\left(\tau_3+i\tau_2\right)\frac{\left(\mathbf{\hat{p}}
-e\mathbf{A}\left(\mathbf{\hat{r}}\right)\right)^2}{2m}
+\tau_3mc^2. \label{f8}
\end{equation}
In this expression $\tau_i$ are the $2\times2$ matrices
\begin{equation}\tau_1=\left(\begin{array}{cc}0&1\\1&0\end{array}\right),
i\tau_2=\left(\begin{array}{cc}0&1\\-1&0\end{array}\right),
\tau_3=\left(\begin{array}{cc}1&0\\0&-1\end{array}\right).
\label{f8a}\end{equation}

Corresponding formalism has been presented by Feshbach and Villars
in Ref. \cite{b10}.

Choose the vector potential in the form:
\begin{equation}
\mathbf{A}\left(\mathbf{r}\right)=\frac{1}{2}\left[\mathbf{B}\times\mathbf{r}\right],
\label{f9}
\end{equation}
where $\mathbf{B}$ has only a $z$ component
\begin{equation}
\mathbf{B}=\left(\begin{array}{c}0\\0\\B\end{array}\right).\label{f10}
\end{equation}
In this case the Hamiltonian (\ref{f8}) can be written as follows:
\begin{equation}
\hat{H}^{mf}=\left(\tau_3+i\tau_2\right)\left(\frac{\hat{p}_z^2}{2m}+
\hbar\omega\left(\hat{n}+\frac{1}{2}\right)\right) +\tau_3mc^2,
\label{f11}
\end{equation}
where
\begin{equation}
\hat{n}=\frac{1}{\hbar\omega}\left(\frac{\hat{p}^2}{2m}+
\frac{m\omega^2}{2}\hat{q}^2\right)-\frac{1}{2}. \label{f12}
\end{equation}
Here $\omega=eB/m$ is the cyclotron frequency. $\hat{p}$ and
$\hat{q}$ are momentum and coordinate after a standard linear
transformation \cite{b14},
\begin{eqnarray}
\hat{q}=\frac{1}{\sqrt{2}}\left(\hat{y}+\frac{\hat{p}_x}{m\omega}\right),
\nonumber\\ \hat{p}=\frac{1}{\sqrt{2}}\left(\hat{p}_y-m\omega
\hat{x}\right).\label{f12a}
\end{eqnarray}
Their physical sense is coordinates of a particle in a frame
connected with the center of the rotational motion (it is clear
that momentum should be redefined in the corresponding units for
such interpretation). One more degree of freedom vanishes in the
Hamiltonian after linear transformation. The physical sense of the
corresponding momentum and coordinate
\begin{eqnarray}
\hat{Q}=\frac{1}{\sqrt{2}}\left(\hat{y}-\frac{\hat{p}_x}{m\omega}\right),
\nonumber\\ \hat{P}=\frac{1}{\sqrt{2}}\left(\hat{p}_y+m\omega
\hat{x}\right)\label{f12b} \end{eqnarray} is of coordinates of the
gyration center.

First of all, consider briefly the well-known example of a
one-dimensional free particle. The Hamiltonian of this system has
the following form:
\begin{equation}
\hat{H}^{fp}=\left(\tau_3+i\tau_2\right)\frac{\hat{p}^2}{2m}
+\tau_3mc^2. \label{f13}
\end{equation}
It is difficult to outline an evident physical meaning for each of
the components of the wave function in this representation (we
denote it as {\em the standard representation}). However, the
matrix of the Hamiltonian (\ref{f13}) can be diagonalized to the
form:
\begin{equation}
\hat{H}^{fp}=\tau_3E(\hat{p}), \label{f14}
\end{equation}
where
\begin{equation}
E(p)=\sqrt{m^2c^4+c^2p^2}. \label{f15}
\end{equation}
One can provide the transformation in this representation {\em(the
Feshbach--Villars representation)} using the following transform
matrix:
\begin{equation}
\hat{U}(\hat{p})\!=\!\frac{1}{2\sqrt{mc^2E(\hat{p})}}\left[\left(E(\hat{p})+mc^2\right)
+\left(E(\hat{p})-mc^2\right)\tau_1\right]. \label{f16}
\end{equation}

The components of the wave function have a sense of the solutions
with positive and negative energies in the Feshbach--Villars
representation.

The coordinate operator in the Feshbach--Villars representation
has the even and odd (diagonal and nondiagonal) parts \cite{b8,
b9, b10}, and can be written in the following form:
\begin{equation}
\hat{q}=i\hbar\partial_p-i\frac{\hbar c^2 p}{2E^2
(p)}\tau_1.\label{f17}
\end{equation}
Such a complicated form of the operator we refer to here as the
nontrivial charge structure of the coordinate operator.

Consider this question closely. It is clear that in our case all
operators are $2\times 2$ operator-valued matrices. In the
Feshbach--Villars representation relevant indices correspond to
different signs of charge (particle or anti-particle). Operators
with a nontrivial charge structure (including operator of
coordinate) are not observables from the point of the charge
superselection rule. Indeed, the measurement of such an operator
shall result in a charge-violating state. However, it is
impossible to build the modern physical theory without such
operators. Most probably the measurement procedure for these
provides a system from a single-particle state to a many-particle
one (i.e., it results in the creation of particle pairs).

On the other hand, the state before measurement is a
single-particle one and is not a superposition of states with
different signs of charge. Hence, the mean of any operator depends
on its even part only (diagonal matrix elements in the charge
space). Therefore, only the even part of the operator is
observable. Detailed consideration of this question has been
presented in the work \cite{b14a}.

One can write the even part of the coordinate operator as follows:
\begin{equation}
\left[\hat{q}\right]=i\hbar\partial_p.\label{f18}
\end{equation}

It is very important that in this case the even part of the
coordinate operator coincides with the Newton--Wigner position
operator $\hat{\xi}$, i.e.,
\begin{equation}
\left[\hat{q}\right]=\hat{\xi}. \label{f19}
\end{equation}
The momentum operator in the Feshbach--Villars representation can
be written in the very simple form
\begin{equation}
\hat{p}=-i\hbar\partial_\xi.\label{f20}
\end{equation}

Now, consider another, more complicated, example -- a relativistic
rotator, i.e, a particle in a constant homogeneous magnetic field
without translational motion along the $z$ axis. In this case, one
can write the Hamiltonian as follows:
\begin{equation}
\hat{H}^{r}=\hbar\omega\left(\tau_3+i\tau_2\right)
\left(\hat{n}+\frac{1}{2}\right) +\tau_3mc^2. \label{f21}
\end{equation}
This Hamiltonian can be written in the Feshbach--Villars
representation (see \cite{b10} and Table \ref{t1}), where it has a
nontrivial charge structure. We use here another representation
\cite{b15}, where the Hamiltonian (\ref{f21}) has a diagonal form.
One can write the transform matrix in this representation as
follows:
\begin{equation}
\hat{U}(\hat{n})\!=\!\frac{1}{2\sqrt{mc^2E(\hat{n})}}\!\left[\left(E(\hat{n})+mc^2\right)
+\left(E(\hat{n})-mc^2\right)\tau_1\right], \label{f22}
\end{equation}
where $E(n)$ is the modulus of the Hamiltonian (\ref{f21})
eigenvalue
\begin{equation}
E(n)=mc^2\sqrt{1+2\lambda^2\left(n+\frac{1}{2}\right)}
,\label{f23}
\end{equation}
and $\lambda=\lambda_c/\sigma$ is the ratio of the Compton
wavelength and the oscillator length. The Hamiltonian (\ref{f21})
in this representation can be written as follows:
\begin{equation}
\hat{H}^{r}=\tau_3E(\hat{n}). \label{f24}
\end{equation}

\begin{table*}[!t]
\caption{\label{t1} The operators for a relativistic rotator in
the different representations. The symbols $\hat{\xi}$ and
$\hat{b}$ correspond to the Newton--Wigner and related
annihilation operators in the column ``Feshbach--Villars
representation'', and to coordinate and annihilation operators of
the nonlocal theory in the column ``Representation of nonlocal
theory.'' Furthermore, the additional symbols used here are as
follows: $S(\hat{p},\hat{n},\hat{p})=
U(\hat{p})R^{\dag}(\hat{n})U^{- 1}(\hat{p})$, $S_\varepsilon
(\hat{p},\hat{n},\hat{p})=U(\hat{p})\varepsilon(\hat{n})R^{\dag}(\hat
{n})U^{-1}(\hat{p})$, $Q(\hat{n},\hat{p},\hat{n}) =
U(\hat{n})i\frac{{\hbar c^2 \hat{p}}}{{2E^2 (\hat{p})}}\tau _1
U^{-1}(\hat{n})$.}
\begin{tabular}{@{}>{\PBS\raggedright\hspace{0pt}}m{0.19\textwidth}@{} @{}>{\PBS\centering\hspace{0pt}}
m{0.27\textwidth}@{}
@{}>{\PBS\centering\hspace{0pt}}m{0.27\textwidth}@{}
@{}>{\PBS\centering\hspace{0pt}}m{0.27\textwidth}@{}}\hline\hline
&Standard representation & Feshbach--Villars representation &
Representation of nonlocal theory
\\ \hline\hline Hamiltonian of relativistic rotator
&$\left({\tau_3+i\tau_2}\right)\hbar\omega\left(\hat{n}+\frac{1}{2}\right)+
mc^2\tau _3$&$\begin{array}{r}\tau_3 E(\hat{p})+\frac{(m \omega
c)^2}{2}\left(\tau _3+i\tau _2\right)\\ \times
E^{-\frac{1}{2}}(\hat{p})\hat{\xi}^2E^{-\frac{1}{2}}(\hat{p})\end{array}$
&$E\left(\hat{n}\right)\tau_3$\\ \hline Standard operator of
coordinate & $q$ & $i\hbar\partial_p-i\frac{\hbar c^2 p}{2E^2
(p)}\tau_1$ & $\begin{array}{l}\frac{1}{2}\hat{\xi}
\left[R(\hat{n})+R(\hat{n}+1)\right]\\
+\frac{i\sigma^2}{2\hbar}\hat{\pi}\left[R(\hat{n})-R(\hat{n}+1)\right]
\end{array}$\\ \hline
Newton--Wigner position operator&$i\hbar\partial_p+i\frac{\hbar
c^2 p}{2E^2 (p)}\tau_1$&$i\hbar\partial_p$ &
$\begin{array}{l}\frac{1}{2}\hat{\xi}
\left[R(\hat{n})+R(\hat{n}+1)\right]\\
+\frac{i\sigma^2}{2\hbar}\hat{\pi}\left[R(\hat{n})-R(\hat{n}+1)\right]
\\ +Q\left(\hat{n},\hat{p},\hat{n}\right)
\end{array}$\\ \hline
Annihilation operator of nonlocal
theory&$\hat{a}R^{\dag}(\hat{n})$&$ \left[\hat{b}-\frac{i\hbar c^2
\hat p}{2\sqrt{2}\sigma
E^2(\hat{p})}\tau_1\right]S(\hat{p},\hat{n},\hat{p})$&$\hat{b}$\\
\hline Even part of standard annihilation operator
&$\hat{a}\varepsilon(\hat{n})R^{\dag}(\hat{n})$&$
\left[\hat{b}-\frac{i\hbar c^2 \hat p}{2\sqrt{2}\sigma
E^2(\hat{p})}\tau_1\right]S_{\varepsilon}(\hat{p},\hat{n},\hat{p})$&$\hat{b}
\varepsilon(\hat{n})$\\ \hline\hline
\end{tabular}
\end{table*}

In this case (under conditions when electric and nonstationary
fields are absent) the Hamiltonian (\ref{f24}) coincides with the
Hamiltonian of the nonlocal theory \cite{b16}. The term
``nonlocal'' has various meanings in contemporary quantum physics.
In this work we use it to note the fact that the Hamiltonian
contains derivatives up to an infinite order. Therefore, this
representation we refer to as {\em the representation of the
nonlocal theory}.

The momentum and coordinate in the operator $\hat{n}$ of the
expression (\ref{f24}) differ from the standard operators
$\hat{p}$ and $\hat{q}$. We call them momentum and coordinate
operators ($\hat{\pi}$ and $\hat{\xi}$) of the nonlocal theory.
These operators with corresponding annihilation and creation
operators
\begin{eqnarray}
\hat{b}=\frac{1}{\sqrt{2}}\left(\frac{\hat{\xi}}{\sigma}+
i\frac{\sigma}{\hbar}\hat{\pi}\right),\label{f25}\\
\hat{b}^{\dag}=\frac{1}{\sqrt{2}}\left(\frac{\hat{\xi}}{\sigma}-
i\frac{\sigma}{\hbar}\hat{\pi}\right)\label{f26}
\end{eqnarray}
have a trivial charge structure.

It is easy to find the evident form of the standard annihilation
operator in the representation of the nonlocal theory:
\begin{equation}
\hat{a}=\hat{b}R\left(\hat{n}\right).\label{f27}
\end{equation}
Here we have introduced the following operator:
\begin{equation}
R(\hat{n})=U\left(\hat{n}-1\right)U^{ - 1}\left(\hat{n}\right)=
\varepsilon\left(\hat{n}\right)+\chi\left(\hat{n}\right)\tau_1.
\label{f28}
\end{equation}
This operator contains both even and odd parts, and the functions
$\varepsilon\left(\hat{n}\right)$ and $\chi\left(\hat{n}\right)$
($\varepsilon$ and $\chi$ factors) are expressed via the energy
spectrum (\ref{f23}):
\begin{equation}
\varepsilon(n)=\frac{{E(n-1)+E(n)}}{{2\sqrt
{E(n-1)E(n)}}},\label{f29}
\end{equation}
\begin{equation}
 \chi(n)=\frac{{E(n-1)-E(n)}}{{2\sqrt{E(n-1)E(n)}}}. \label{f30}
\end{equation}

Hence, the even part of the annihilation operator can be written
in the following form:
\begin{equation}
\left[\hat{a}\right]=\hat{b}\varepsilon\left(\hat{n}\right).\label{f31}
\end{equation}
It should be noticed that unlike the case of a free particle, the
even part of the standard coordinate (annihilation) operator does
not coincide with the corresponding operator of the nonlocal
theory. Furthermore, the even part of the annihilation operator is
a deformed annihilation operator of the nonlocal theory (similar
to Eq. (\ref{f6})), where the $\varepsilon$ factor is the
deforming function.

Similar to \cite{b8, b9, b10} and taking into account the above
discussion, the even part of the standard coordinate and momentum
(creation and annihilation) operators play the role of ``mean
positions'', and are a deformed canonical pair here. Their
commutator can be written in the following form:
\begin{equation}
\left[{[\hat{a}],[\hat{a}]^{\dag}}\right]=
\varepsilon^{2}\left(\hat{n}+1\right)\left(\hat{n}+1\right)-
\varepsilon^{2}\left(\hat{n}\right)\hat{n}.\label{f32}
\end{equation}
The reason for this deformation is the interaction of a particle
with the vacuum. Though particle pair creation is absent here, a
particle feels the vacuum structure, and this is described by the
$\varepsilon$ factor.

This peculiarity leads to one more unexpected feature. Consider
even parts of the coordinate operators $\hat{x}$ and $\hat{y}$ of
the usual configuration space. Using Eq. (\ref{f12a}) and
(\ref{f12b}) one can obtain the relevant commutation relation:
\begin{equation}
\left[{[\hat{x}],[\hat{y}]}\right]=\frac{i\sigma^2}{2}\left(
\varepsilon^{2}\left(\hat{n}+1\right)\left(\hat{n}+1\right)-
\varepsilon^{2}\left(\hat{n}\right)\hat{n}-1\right),\label{f32a}
\end{equation}
where $\sigma$ is the oscillator length. Hence, the ``mean
position'' operators for the standard configuration space do not
commute with each other. This fact distinguishes between the
relativistic theory and the nonrelativistic (and nonlocal) one and
is a consequence of the vacuum influence as well. Note that the
nature of this noncommutativity differs from one presented in the
works \cite{b16a,b16b}.

The last example that we consider here is the particle in a
constant homogeneous magnetic field, i.e., the relativistic
rotator with translational motion along the $z$ axis. This system
is described by the Hamiltonian (\ref{f11}). The modulus of an
eigenvalue has the following form:
\begin{equation}
E(n,p_z)=mc^2\sqrt{1+2\lambda^2\left(n+\frac{1}{2}\right)+\frac{p_z^2}{m^2c^2}},
\label{f33}
\end{equation}

In this case, the standard operators in the representation  of the
nonlocal theory can be written as follows:
\begin{equation}
\hat{z}=i\hbar\partial_{p_z}-i\frac{\hbar c^2 p_{z}}{2E^2
(\hat{n},p_z)}\tau_1,\label{f34}
\end{equation}
\begin{equation}
\hat{a}=\hat{b}R\left(\hat{n},p_z\right),\label{f35}
\end{equation}
where the operator $R\left(\hat{n},p_z\right)$, with $\varepsilon$
and $\chi$ factors determined similarly to the expressions
(\ref{f28}), (\ref{f29}), (\ref{f30}). Therefore, the even parts
of these operators have the following form:
\begin{equation}
\left[\hat{z}\right]=i\hbar\partial_{p_z},\label{f36}
\end{equation}
\begin{equation}
\left[\hat{a}\right]=\hat{b}\varepsilon\left(\hat{n},p_z\right).\label{f37}
\end{equation}
A specific peculiarity of these expressions is the fact that the
``mean positions'' operators of the rotational and translational
motions do not commute with each other. The corresponding
commutator is written as follows:
\begin{equation}
\left[{[\hat{z}],[\hat{a}]}\right]=ip_z\hat{b}\frac{\hbar
m^2c^6}{2}
\frac{E\left(\hat{n},p_z\right)-E\left(\hat{n}-1,p_z\right)}
{\left(E\left(\hat{n},p_z\right)E\left(\hat{n}-1,p_z\right)\right)^{\frac{5}{2}}}
.\label{f38}
\end{equation}
It means that not only the configuration space of the rotational
motion is noncommutative, but the $z$ ``mean position'' does not
commute with those degrees of freedom due to the properties of the
vacuum.

The coherent states that satisfy both conditions presented in the
Introduction, can be constructed as eigenstates of the even part
of the annihilation operator. The properties of such states are
considered in the following sections.

\section{Coherent states for a free particle}
\label{s3}

According to Eqs. (\ref{f19}) and (\ref{f20}), the even part of
the annihilation operator is the annihilation operator of the
nonlocal theory in the case of a free particle:
\begin{equation}
[\hat{a}]=\hat{b}=\frac{1}{\sqrt{2}}\left(\frac{\hat{\xi}}{\sigma}+i
\frac{\sigma}{\hbar}\hat{p}\right).\label{f39}
\end{equation}
It is clear that $\sigma$ is not the oscillator length but an
arbitrary parameter. Hence, the coherent state of a relativistic
free particle is the standard coherent state with one component of
charge. In the Feshbach--Villars representation this state can be
written in the following form:
\begin{equation}
\Psi_{\alpha,\pm}\!\left(p\right)\!=\!\frac{1}{2\pi^{1/4}}\!
\left(\!\begin{array}{c}1\!\pm\!
1\\1\!\mp\!1\end{array}\!\right)\!\exp\!\left(\!-\!\frac{1}{2}\!
\left(\!p\!-\!\sqrt{2}\alpha^{\prime\prime}\!
\right)^{2}\!-\!i\!\sqrt{2}\alpha^{\prime}\!p\!\right)\!.
\label{f40}
\end{equation}
Here we have used the dimensionless units, where the momentum is
measured in units of $\hbar/\sigma$ and the coordinate is measured
in units of $\sigma$.

\begin{figure}[h]
\begin{center}
\includegraphics[width=\columnwidth,clip=]{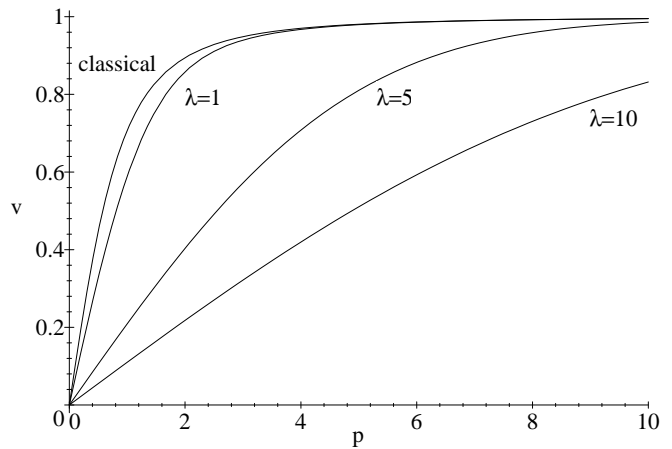}
\end{center}
\caption{\label{fig1} Dependence of the mean velocity on the mean
momentum for the coherent states of a free particle with different
values of the parameter $\lambda$ and for the classical
(nonquantum) case. Momentum is given in $m c$ units; velocity is
given in the speed of light units.}
\end{figure}

The trajectory of the wave packet in this case is the same as in
the nonlocal theory. The mean velocity (in the speed of light
units) can be written as the following integral:
\begin{equation}
\bar{v}=\frac{\lambda}{\sqrt{\pi}}\int\limits_{-\infty}^{+\infty}
\frac{p}{\sqrt{1+\lambda^{2}p^{2}}}\exp\left(-\left(p-\sqrt{2}
\alpha^{\prime\prime}\right)^{2}\right)dp .\label{f41}
\end{equation}
This expression can be rewritten in the form of the power series:
\begin{equation}
\bar{v}\!=\!\frac{\lambda\!\sqrt{2}\alpha^{\prime\prime}}{\sqrt{\pi}}\!
\sum\limits_{n=0}^{\infty}\!\left(\!\sqrt{2}\alpha^{\prime\prime}\!\lambda\!\right)^{2n}\!
\sum\limits_{k=0}^{\infty}\!\lambda^{2k}\!C_{-1/2}^{n+k}\!C_{2(n+k)+1}^{2k}\!
\Gamma\!\left(\!k\!+\!\frac{1}{2}\!\right)\!.\label{f42}
\end{equation}
The parameter $\lambda$ that describes the localization of the
wave packet is the independent relativistic parameter along with
mean momentum. In Fig.~\ref{fig1} we plot the dependence of the
mean velocity on the mean momentum for different values of
$\lambda$ and for the classical (nonquantum) case.

From this plot one can see that for small values of the momentum,
this dependence is a nonrelativistic one but with larger mass.
Indeed, writing the power series (\ref{f42}) with terms $n=0$
only, one can obtain the following expression (in the dimensional
units):
\begin{equation}
\bar{v}=\frac{\bar{p}}{m^{*}}, \label{f43}
\end{equation}
where we have introduced the effective mass as follows:
\begin{equation}
m^{*}=\frac{m}{\frac{2}{\sqrt{\pi}}\sum\limits_{n=0}^{\infty}
\lambda^{2n}C^{n}_{-1/2}\Gamma\left(n+\frac{3}{2}\right)}.\label{f44}
\end{equation}

The difference between the standard and nonlocal theories can be
verified on the expected values of observables, which contain
second and higher powers of the coordinate. We consider it on the
example of the dispersion of coordinate.

The second moment of the coordinate (\ref{f17}) contains both the
standard and additional terms. The reason for this is the fact
that the square of an odd operator is an even operator. Hence, the
square of the dispersion for the coherent state (\ref{f40}) can be
written (in dimensionless units) as follows:
\begin{equation}
\overline{\Delta
q^2}\!=\!\frac{1}{2}-\frac{\lambda^4}{4\sqrt{\pi}}\!\int\limits_{-\infty}^{+\infty}\!
\frac{p^2}{\left(1+\lambda^2p^2\right)^2}\exp\left(-\left(p-\sqrt{2}\alpha^
{\prime\prime}\right)^2\right)d\!p\!.\label{f45}
\end{equation}

\begin{figure}[ht]
\begin{center}
\includegraphics[width=\columnwidth,clip=]{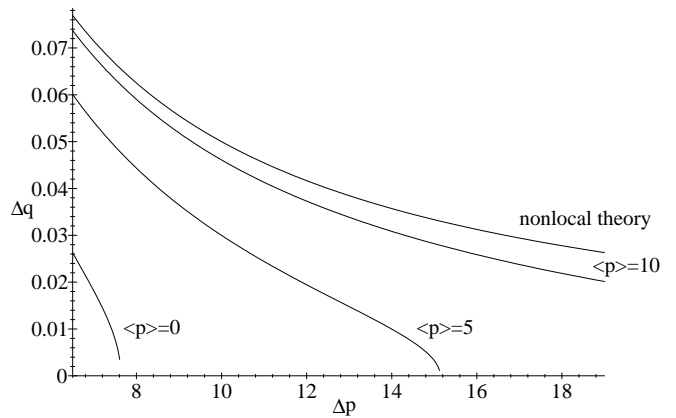}
\end{center}
\caption{\label{fig2} Dependence of the coordinate dispersion on
the momentum dispersion for coherent states of a free particle
with different values of the mean momentum $\left\langle
p\right\rangle=\sqrt{2}\alpha^{\prime\prime}\lambda$. Dispersion
of the coordinate is given in $\lambda_c=\hbar/mc$ units;
dispersion of the momentum and mean momentum are given in $mc$
units. For comparison, the curve $\Delta q=1/(2\Delta p)$ for the
coherent states in the nonlocal theory is given.}
\end{figure}

In Fig.~\ref{fig2} we plot the dependence of the coordinate
dispersion on the momentum dispersion for coherent states with
different mean momenta. It should be noticed that there exists a
formal violation of the uncertainty relation for all these states,
especially for the large dispersion of momentum. For the very
localized states, the square of dispersion even can be negative.
This is a specific feature of the spin-$0$ particles, which are
described by the Klein--Gordon equation. The reason for it is the
indefinite metric of the Hilbert space of states. This fact makes
for difficulties in the probability interpretation of these
particles.

These peculiarities, which lead from the nontrivial charge
structure of the coordinate operator, vanish for the strong
localized states with a large mean momentum. It means that, in
this case, the approximation of the nonlocal theory can be used
for a description of scalar charge particles. Nevertheless, for
strongly localized states with a small mean momentum these
peculiarities manifest themselves.

\section{Coherent states for relativistic rotator}
\label{s4}

The even part (\ref{f31}) of the standard annihilation operator is
a deformed annihilation operator of the nonlocal theory in the
case of a relativistic rotator. Hence, the corresponding coherent
states are the so-called nonlinear coherent states \cite{b6, b7}
with the $\varepsilon$ factor as a deforming function:
\begin{equation}
\left|\alpha,\pm\right\rangle=\mathcal{N}^{-\frac{1}{2}}
\left(\left|\alpha\right|^2\right)\sum
\limits_{n=0}^{\infty}\frac{\alpha^{n}}{\sqrt{n!}\left[\varepsilon(n)\right]!}
\left|n\right\rangle\otimes\left|\pm\right\rangle.\label{f46}
\end{equation}
Here $\left|\pm\right\rangle$ is the charge part of the quantum
state. The functional factorial and the normalization factor are
determined in the usual way:
\begin{equation}
\left[\varepsilon(n)\right]!=\left\{ \begin{array}{ll} 1 &
\textrm{if $n=0$}\\ \prod\limits_{k=1}^{n}\varepsilon(k) &
\textrm{if $n=1,\ldots,\infty$}\end{array}\right.,\label{f47}
\end{equation}
\begin{equation}
\mathcal{N}\left(\left|\alpha\right|^2\right)=\sum
\limits_{n=0}^{\infty}\frac{\left|\alpha\right|^{2n}}{n!\left[\varepsilon^2(n)\right]!}
.\label{f48}
\end{equation}

The $\varepsilon$ factor is the difference between the coherent
states in the standard and nonlocal theories. It results in the
fact that not only the dispersions of coordinate and momentum have
some peculiarities, but the expected value of the coordinate and
momentum as well.

Consider briefly the properties of the function
$\left[\varepsilon(n)\right]!$. From the definition (\ref{f29}),
the $\varepsilon$ factor
\begin{equation}
\varepsilon(n)
=1+\frac{5\lambda^4+3}{128\lambda^4}\frac{1}{n^2}+O\left(\frac{1}{n^3}\right),
\label{f49}
\end{equation}
for large $n$. Therefore the quantity $[\varepsilon(n)]!$
converges to a nonzero, finite factor as   $n\rightarrow\infty$.
For example, it is possible to find nonzero numbers $a$ and $b$
such that
\begin{equation}
\exp\left(\frac{a}{n^2}\right)<\varepsilon\left(n\right)<
\exp\left(\frac{b}{n^2}\right)\label{f50}
\end{equation}
holds for all $n>0$. In this case
\begin{equation}
\exp\left(\frac{\pi^2a}{6}\right)<\lim_{n\rightarrow\infty}
\left[\varepsilon(n)\right]!<\exp\left(\frac{\pi^2b}{6}\right).\label{f51}
\end{equation}
Hence this case has the property that the terms with the large $n$
are effectively of the canonical form where $\sqrt{n!}$ is the
controlling factor. From this fact we would be certainly inclined
to believe that $\varepsilon$ factor influences on the particle
behavior are very low.

Consider this fact on the example of the expected values of the
coordinate and momentum time evolution. The Heisenberg equation in
the representation of the nonlocal theory has the following form:
\begin{eqnarray}
\partial_t[\hat{a}]=-\frac{i}{\hbar}\tau_3\left(E\left(\hat{n}+1\right)-
E\left(\hat{n}\right)\right)[\hat{a}],\label{f52}\\
\partial_t\{\hat{a}\}=\frac{i}{\hbar}\tau_3\left(E\left(\hat{n}+1\right)+
E\left(\hat{n}\right)\right)\{\hat{a}\} ,\label{f53}
\end{eqnarray}
where $[\hat{a}]$ and $\{\hat{a}\}$ are the even and odd parts of
the standard annihilation operator. Their solution can be written
as follows:
\begin{eqnarray}
[\hat{a}]\left(t\right)=\exp\left(-\frac{i}{\hbar}\tau_3\left(E\left(\hat{n}+1\right)-
E\left(\hat{n}\right)\right)t\right)[\hat{a}],\label{f54}\\
\{\hat{a}\}\left(t\right)=\exp\left(\frac{i}{\hbar}\tau_3\left(E\left(\hat{n}+1\right)+
E\left(\hat{n}\right)\right)t\right)\{\hat{a}\}
 .\label{f55}
\end{eqnarray}
The expected value of this operator in the coherent state
(\ref{f46}) is
\begin{equation}
\bar{a}\!\left(t\right)\!=\!\pm\!
\alpha\!\mathcal{N}^{-1}\!\left(\!\left|\alpha\right|^2\!\right)\!\sum
\limits_{n=0}^{\infty}\!\frac{\left|\alpha\right|^{2n}}{n\!!\!\left[\!\varepsilon^2(n)\!\right]\!!}\!
\exp\!\left(\!\mp\!\frac{i}{\hbar}\!\left(\!E\!\left(\!n\!+\!1\!\right)\!-\!
E\!\left(\!n\!\right)\!\right)\!t\!\right)\!.\label{f56}
\end{equation}

In the first relativistic correction, the dependence on the
$\varepsilon$ factor vanishes and this expected value can be
calculated analytically \cite{b13}: \setlength\arraycolsep{2pt}
\begin{eqnarray}
\bar{a}\left(t\right)&=&\pm\!
\alpha\exp\left(-2\left|\alpha\right|^2\sin^2\left(\frac{\lambda^2\omega
t }{2}\right)\right)\nonumber \\ &\times&\!\exp\left(\mp
i\left[\left(1-\lambda^2\right)\omega
t-\left|\alpha\right|^2\sin\left(\lambda^2\omega
t\right)\right]\right)\! . \label{f57}
\end{eqnarray}
Along with the cyclotron frequency, there exists the low frequency
\begin{equation}
\Omega=\lambda^2\omega<\omega,\label{f58}
\end{equation}
that modulates the rotational motion. However, due to the
bremsstrahlung, it can be regarded as an additional damping rather
than a low frequency modulation in the real physical systems.

\begin{figure*}[t]
\begin{center}
\mbox{\includegraphics[width=0.45\textwidth,clip=]{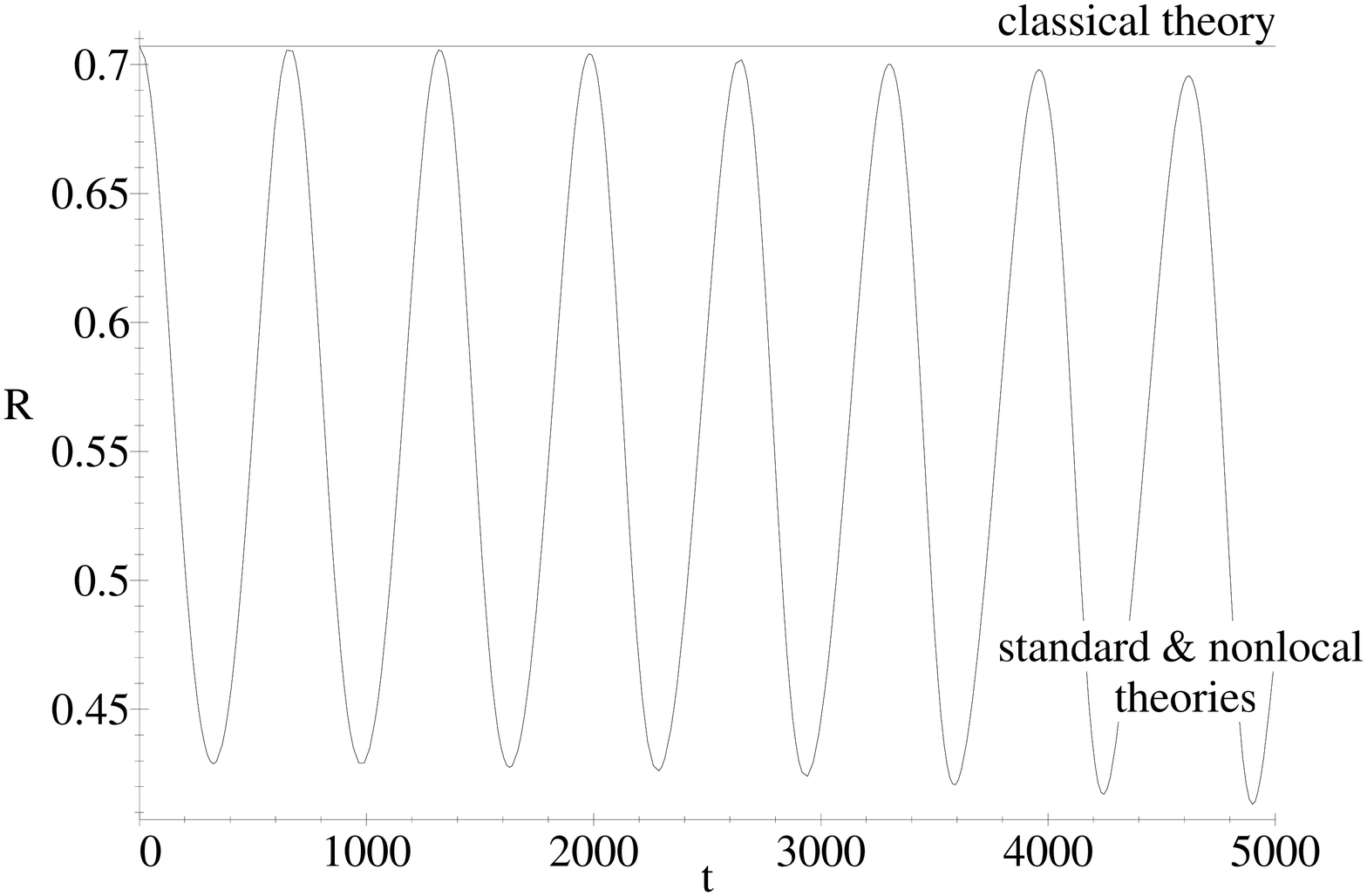}{\it
(a) }\includegraphics[width=0.45\textwidth,clip=]{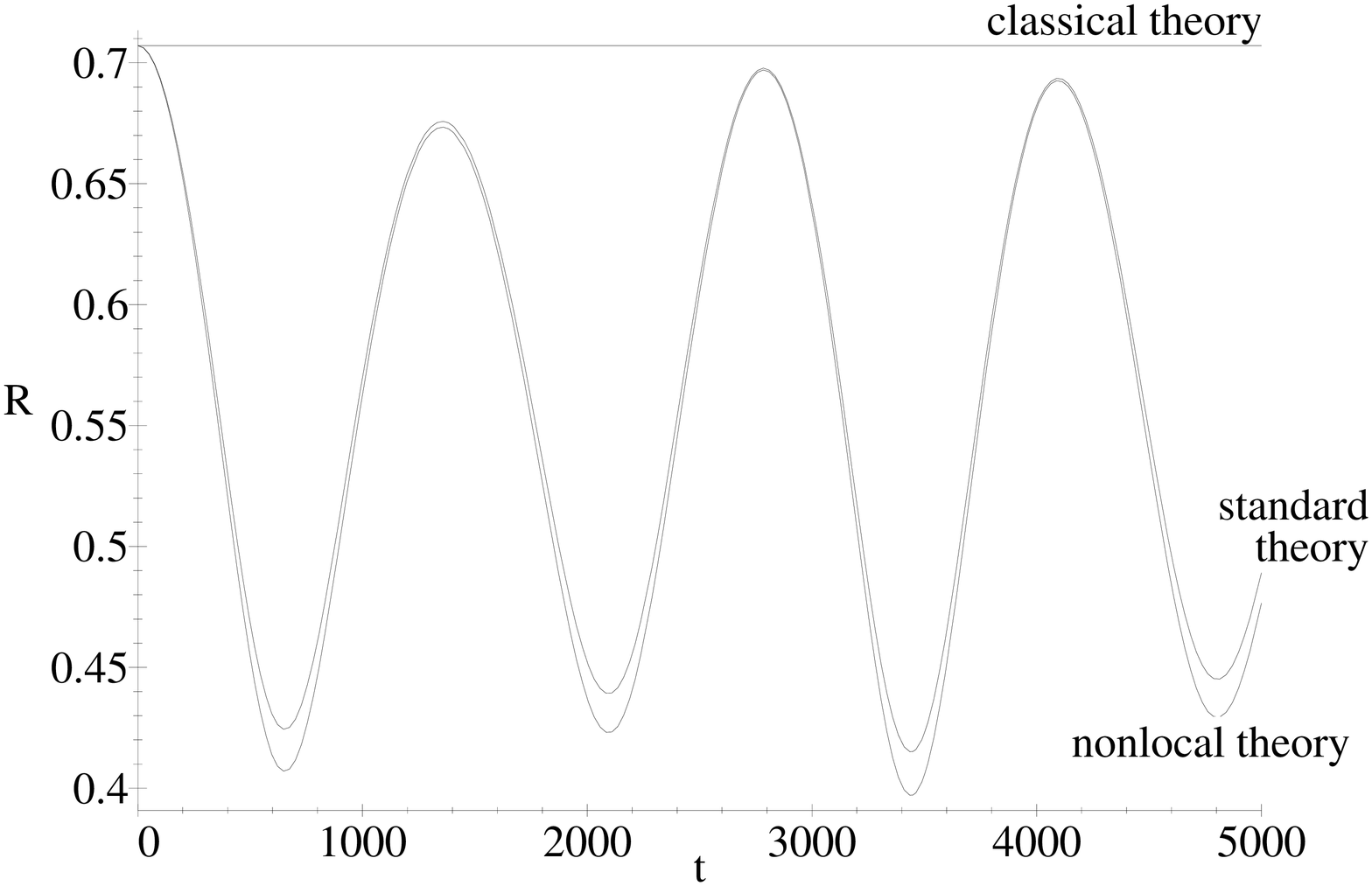}{\it
(b) }}
\mbox{\includegraphics[width=0.45\textwidth,clip=]{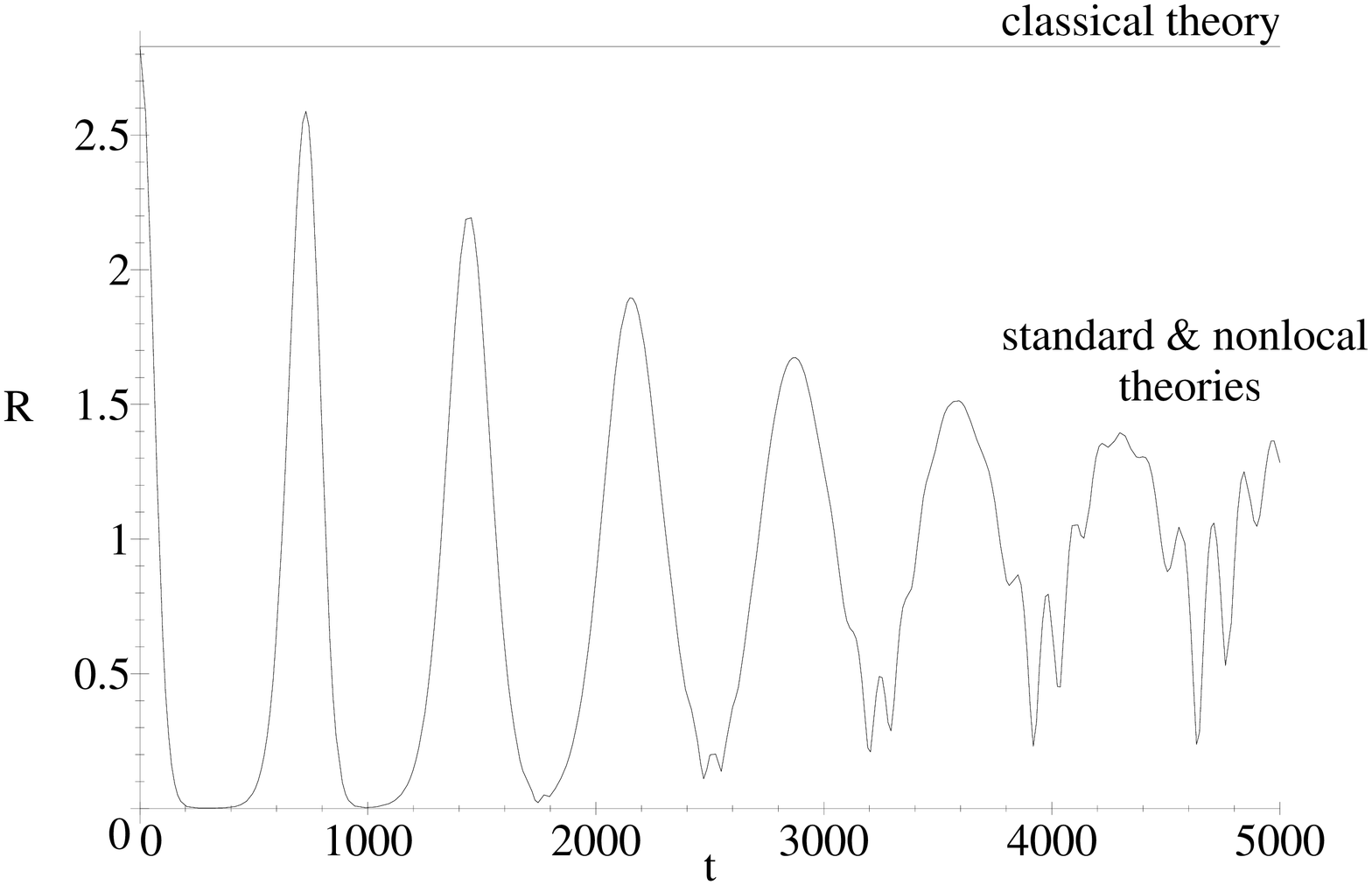}{\it
(c) }\includegraphics[width=0.45\textwidth,clip=]{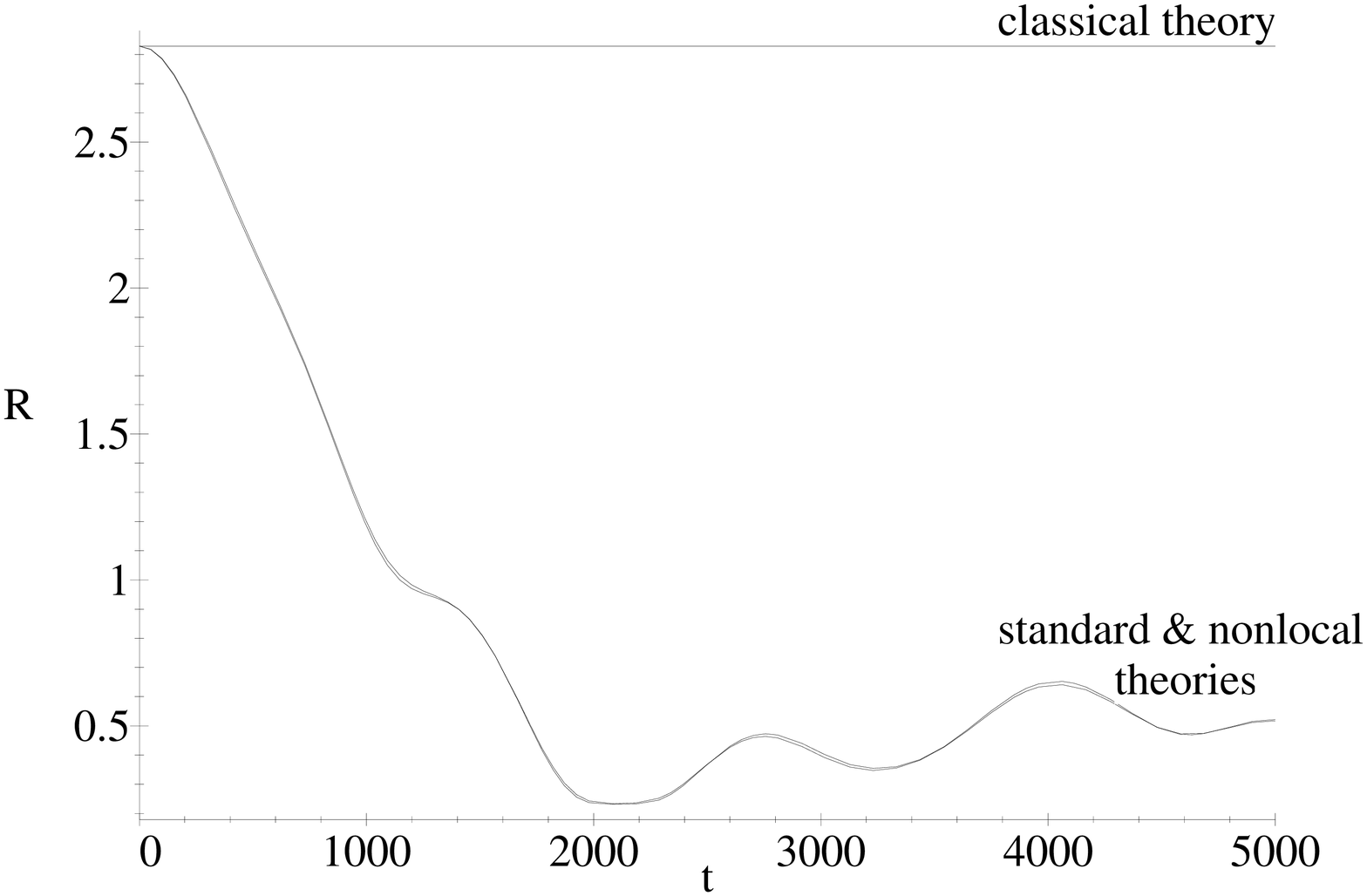}{\it
(d) }}
\end{center}
\caption{\label{fig3} The time evolution of the mean gyration
radius of a relativistic rotator in classical and quantum
relativistic (standard and nonlocal) theories. The radius is given
in the dimensionless units; time is given in $1/\omega$ units.
{\it (a) } $\lambda=0.1$, $\overline{q}=0.5$, $\overline{p}=0.5$.
This case can be calculated analytically. {\it (b)} $\lambda=50$,
$\overline{q}=0.5$, $\overline{p}=0.5$. In this case one can
distinguish between the standard and nonlocal theories. {\it (c)}
$\lambda=0.1$, $\overline{q}=2$, $\overline{p}=2$. {\it (d)}
$\lambda=50$, $\overline{q}=2$, $\overline{p}=2$. In the all cases
(not only with small $\lambda$) the low-frequency harmonic
exists.}
\end{figure*}

In Fig.~\ref{fig3} we plot the results of numerical calculations
for the time evolution of the mean gyration radius. First of all
it should be noticed that dependence on the $\varepsilon$-factor
is very small. Only in the case where $\lambda$ is large enough
and the initial radius small, one can distinguish between the
standard and nonlocal theories. This difference is very
insignificant. Also, we note that the low-frequency is intrinsic
to all cases. However, in real physical systems this effect should
be very suppressed by the bremsstrahlung.

Similar to the case of a free particle, the more significant
difference between the standard and nonlocal theories can be
exhibited in the second moments. Consider the square of the
gyration radius
\begin{equation}
\hat{R}^2=2\hat{n}+1.\label{f59}
\end{equation}
Note that this observable is an integral of motion. Then, one can
find the dispersion of the gyration radius as follows:
\begin{equation}
\overline{\Delta
R^2}=1-\overline{R}^2\left(1-\mathcal{N}^{-1}\left(\overline{R}^2/2\right)
\sum\limits_{n=0}^{\infty}\frac{\overline{R}^{2n}}
{2^{n}n!\left[\varepsilon^2(n+1)\right]!}\right),\label{f60}
\end{equation}
where $\overline{R}^{2}=2\left|\alpha\right|^2$.

\begin{figure}[h]
\begin{center}
\includegraphics[width=\columnwidth,clip=]{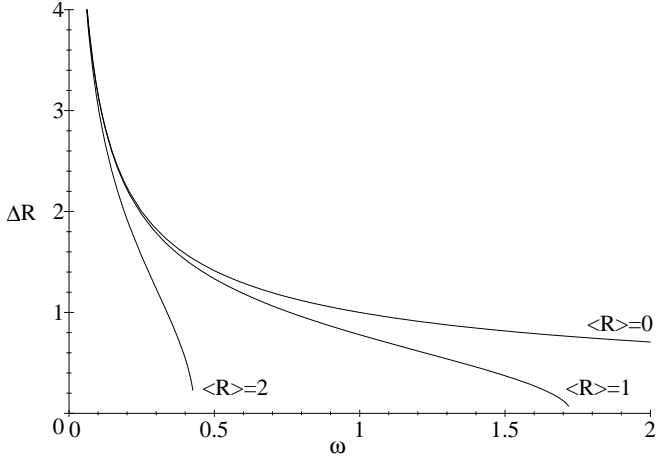}
\end{center}
\caption{\label{fig4} Dependence of the gyration radius dispersion
on the cyclotron frequency for coherent states of a relativistic
rotator with different values of the mean gyration radius
$\left\langle R\right\rangle$. The cyclotron frequency is given in
$mc^2/\hbar$ units; the dispersion of gyration radius and mean
radius are given in $\lambda_c=\hbar/mc$ units. Case $\left\langle
R\right\rangle=0$ coincides with a curve for a coherent state in
the nonlocal theory.}
\end{figure}

The results of the numerical calculations for the dependence of
the gyration radius dispersion on the cyclotron frequency are
given in the Fig.~\ref{fig4}. Unlike the case of a free particle,
the peculiarities resulting from the nontrivial charge structure
of the coordinate and momentum operators are more evident for the
coherent states with large parameter $\alpha$ (mean gyration
radius). The case of $\alpha=0$ corresponds to the eigenstate of
the Hamiltonian with $n=0$. All peculiarities are absent here.

\section{Coherent states for a particle in a constant homogeneous
magnetic field}
\label{s5}

The consideration of a particle in a constant homogeneous magnetic
field combines the two previous cases. However, it is impossible
to construct coherent states that satisfy our conditions for both
the translational and rotational degrees of freedom
simultaneously, because the even parts of the corresponding
annihilation operators do not commutate with each other (see Eq.
(\ref{f38})). Therefore we will define coherent states for each
degree of freedom separately.

The coherent state for the translational motion has the following
form:
\begin{equation}
\left|n,\alpha_z,\pm\right\rangle=\left|n\right\rangle
\otimes\left|\alpha_z\right\rangle\otimes\left|\pm\right\rangle,\label{f61}
\end{equation}
where $\left|\alpha_z\right\rangle$ is the standard coherent state
for the $z$ axis and $\left|n\right\rangle$ is the eigenstate of
the rotator. The coherent state for the rotational motion can be
written as follows:
\begin{equation}
\left|\alpha_r,p_z,\pm\right\rangle\!=\!\mathcal{N}^{-\frac{1}{2}}
\left(\!\left|\alpha_r\right|^2,\!p_z\!\right)\!\sum
\limits_{n=0}^{\infty}\!\frac{\alpha_r^{n}}{\sqrt{n!}\!\left[\!\varepsilon(\!n,p_z\!)\!\right]\!!}
\!\left|n\right\rangle\!\otimes\!\left|p_z\right\rangle\!\otimes\!
\left|\pm\right\rangle\!,\label{f62}
\end{equation}
where $\left|p_z\right\rangle$ is the eigenstate of the $z$
component of momentum and the normalization factor is determined
in the form:
\begin{equation}
\mathcal{N}\left(\left|\alpha_r\right|^2,p_z\right)=\sum
\limits_{n=0}^{\infty}\frac{\left|\alpha_r\right|^{2n}}{n!\left[\varepsilon^2(n,p_z)\right]!}
.\label{f63}
\end{equation}
It should be noticed that in the present form the coherent state
(\ref{f62}) includes an eigenstate of the $z$ component of
momentum. Hence, it is normalized on $\delta$ function
\begin{equation}
\left\langle\pm,p^{\prime}_z,\alpha_r|\alpha_r,p_z,\pm\right\rangle
=\pm\delta\left(p_z-p^{\prime}_z\right). \label{f63a}
\end{equation}

Generally speaking, the states (\ref{f61}) and (\ref{f62}) are
direct products of coherent states on the one degree of freedom
and part of the Hamiltonian eigenstates on another one. These
states can be redefined in such a way to be the coherent states
for both degrees of freedom. However, they will not satisfy
condition \ref{i1} from the Introduction for one of them.

Summing the states (\ref{f61}) on $n$ with
$e^{-\frac{\left|\alpha_r\right|^2}{2}}\alpha_r^n/\sqrt{n!}$, one
can obtain the coherent states of the nonlocal theory presented in
\cite{b12},
\begin{equation}
\left|\alpha_r,\alpha_z,\pm\right\rangle=\left|\alpha_r\right\rangle
\otimes\left|\alpha_z\right\rangle\otimes\left|\pm\right\rangle,\label{f64}
\end{equation}
where $\left|\alpha_r\right\rangle$ is a standard coherent state
for the rotational degree of freedom. For the rotational motion
these states do not satisfy condition \ref{i1} from the
Introduction.

Now, integrating the states (\ref{f62}) with
$\Psi_{\alpha_z}\left(p_z\right)$ (the standard coherent states in
the momentum representation), one can obtain the following
coherent states:
\setlength\arraycolsep{1pt}
\begin{eqnarray}
\left|\alpha_r,\alpha_z,\pm\right\rangle && \nonumber \\ \!=\sum
\limits_{n=0}^{\infty}\!\int\limits_{-\infty}^{\infty}\!d&\!p_z\!&
\!\Psi_{\alpha_z}\!\left(\!p_z\!\right)\!
\mathcal{N}\!^{-\frac{1}{2}}\!
\left(\!\left|\!\alpha_r\!\right|^2\!,\!p_z\!\right)\!\frac{\alpha_r^{n}}
{\sqrt{n!}\!\left[\!\varepsilon\!(\!n\!,\!p_z\!)\!\right]\!!}\!
\left|\!n\!\right\rangle\!\otimes\!\left|\!p_z\!\right\rangle\!\otimes\!
\left|\!\pm\!\right\rangle\!.\nonumber \\ \label{f65}
\end{eqnarray}
These coherent states do not satisfy condition \ref{i1} for the
translational motion along the field \cite{b13}.
Nevertheless, they are eigenstates of the even part of the
annihilation operator (\ref{f37}), and describe the rotational
motion taking into account a finite localization along the $z$
axis.

In the approximation of the nonlocal theory and for the first
relativistic correction of the standard theory, the states
(\ref{f64}) and (\ref{f65}) are the same. Hence, one can find the
time evolution of the first moments of the rotational motion in
this case as follows:
\begin{widetext}
\begin{eqnarray}
\overline{a_r}\left(t\right)&=&\pm\alpha_r
\frac{1}{\sqrt[4]{1+\frac{\lambda_z^4\omega^2t^2}{4}}}
\exp\left(-2\left|\alpha_r\right|^2\sin^2\left(\frac{\lambda_r^2\omega
t }{2}\right)-\frac{\alpha_z^{\prime\prime
2}\lambda_z^4\omega^2t^2}{2+\frac{\lambda_z^4\omega^2t^2}{2}}\right)\nonumber
\\
&\times&\exp\left(\mp i\left[\left(1-\lambda_r^2\right)\omega
t-\left|\alpha_r\right|^2\sin\left(\lambda_r^2\omega t\right)
-\frac{\alpha_z^{\prime\prime
2}\lambda_z^2\omega^2t^2}{1+\frac{\lambda_z^4\omega^2t^2}{4}}+
\frac{1}{2}\arctan\left(\frac{\lambda_z^4\omega^2t^2}{4}\right)\right]\right).\label{f66}
\end{eqnarray}
\end{widetext}
Therefore, from the comparison of this equation and Eq.
(\ref{f57}), one can conclude that the localization along the $z$
axis leads to the peculiarities of the rotational motion for the
time
\begin{equation}
t>\frac{1}{\lambda_z\omega}.\label{f67}
\end{equation}

For the mean velocity of the translational motion along $z$ axis
we have the following:
\begin{equation}
\bar{v}_z=\overline{p}_z\left[\frac{1}{m^*}-\frac{\hbar\omega}{m^2c^2}
\left(\left|\alpha_r\right|^2+\frac{1}{2}\right)
\right].\label{f68}
\end{equation}
The second term in this equation is the same in the classical
(nonquantum) theory. Hence, the peculiarities of the translational
motion result only from the localization along this degree of
freedom in such approximation.

\section{Coherent states in the Wigner representation}
\label{s6}

The localization peculiarities, which have been described in the
preceding sections, can be illustrated by means of the
relativistic Wigner function for charge-invariant observables
\cite{b17, b18}. The nontrivial charge structure of the coordinate
and momentum operators is taken into account in this
representation.

Utilizing the definition of the Wigner function from \cite{b17},
one can obtain for the coherent state of a free particle
(\ref{f40}) (for determination, we consider only positive charge):
\setlength\arraycolsep{2pt}
\begin{eqnarray}
&&W_{[+],\alpha}\!(p,q)\!=\!\frac{1}{\pi^{3/2}}\!\exp\!\left(\!-\left(p-
\sqrt{2}\alpha^{\prime\prime}\right)^2\!\right) \nonumber \\
&&\!\times\!\int\limits_{-\infty}^{+\infty}\!
\left(\!\frac{1\!+\!\lambda^2\!\left(\!p\!+\!x\!\right)^2}{1\!+\!\lambda^2\!
\left(\!p\!-\!x\!\right)^2}\!\right)^{1/4}\!\exp\!\left(\!-x^2\!\right)\!\cos\!
\left(\!2\!\left(\!q\!-\!\sqrt{2}\alpha^{\prime}\!\right)\!x\!\right)\!d\!
x\!.\nonumber
\\ \label{f69}
\end{eqnarray}

\begin{figure}[t]
\begin{center}
\mbox{\includegraphics[width=\columnwidth-1.7em,clip=]{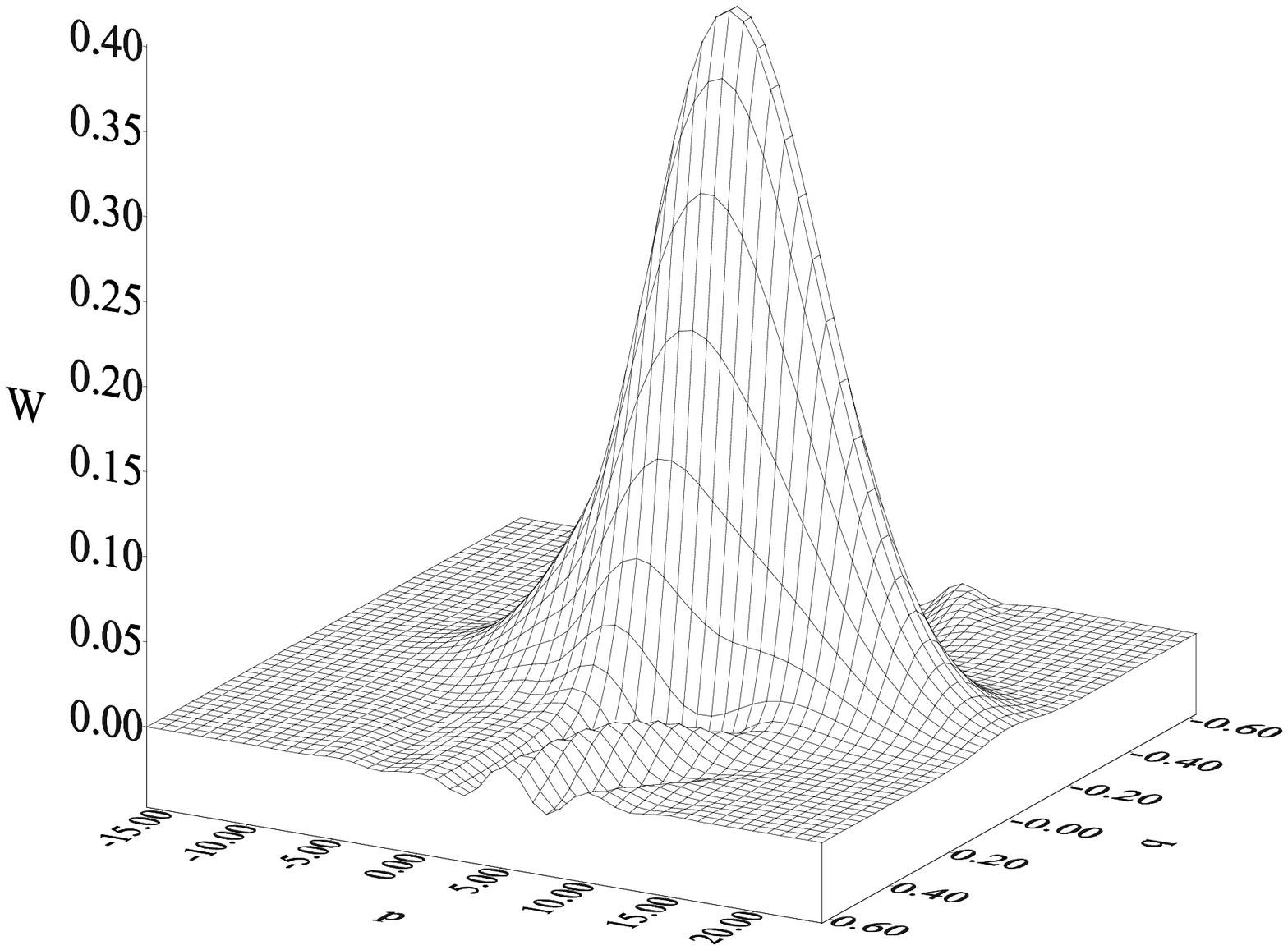}{\it
(a) }}
\mbox{\includegraphics[width=\columnwidth-1.7em,clip=]{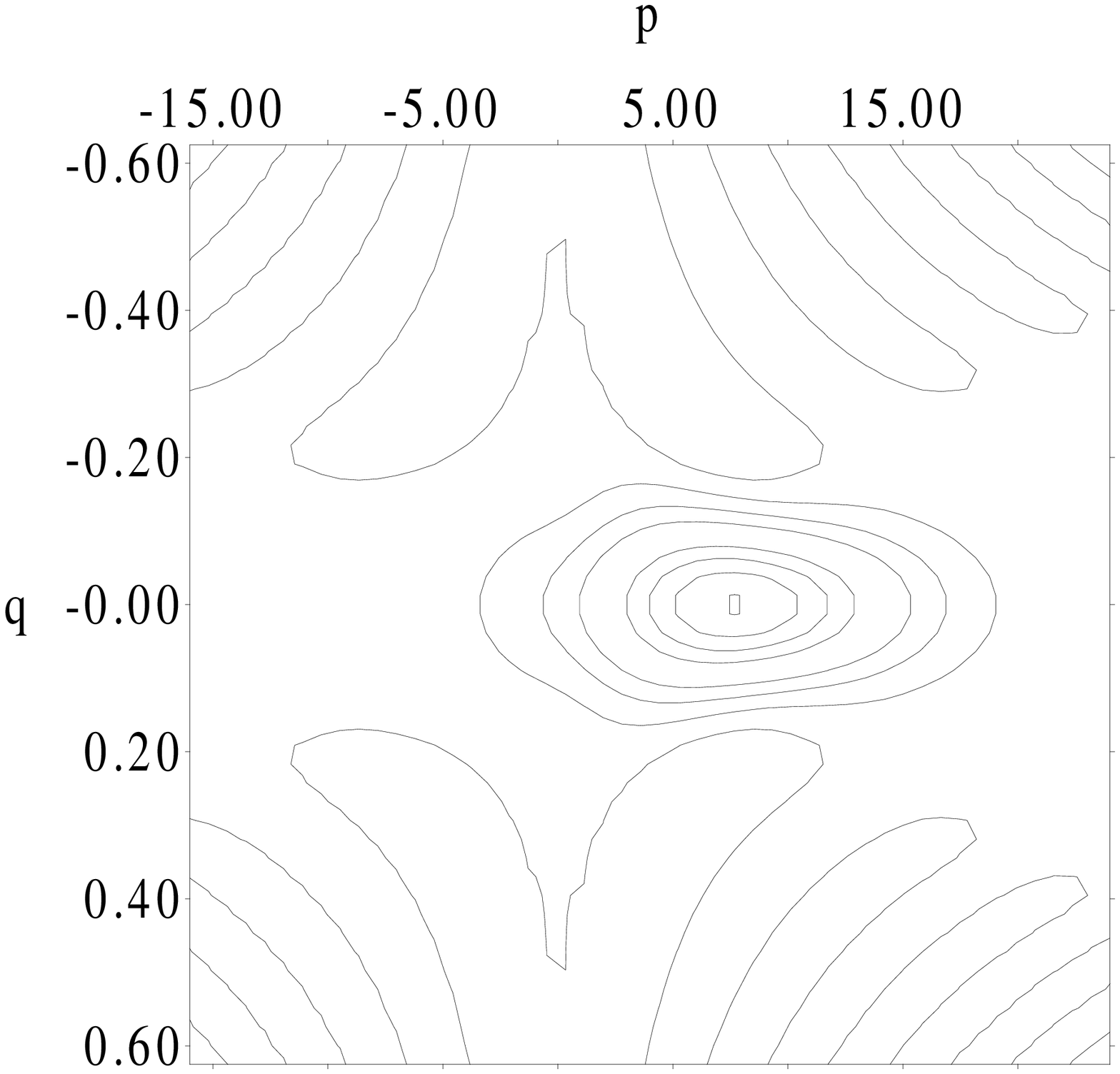}{\it
(b)}}
\end{center}
\caption{\label{fig5} The Wigner function {\it (a) } and its
contours {\it (b)} for the coherent state of a free particle. The
ratio of the Compton wavelength and character wave-packet length
is $\lambda=8$. The mean momentum is $8mc$. In this plot momentum
is given in $m c$ and the coordinate in $\lambda_c=\hbar/m c$
units.}
\end{figure}

Due to the negative square of the coordinate dispersion for strong
localized states, there exists the supposition that such states
are impossible. In Fig.~\ref{fig5} we plot the Wigner function for
such a state. One can see that negative square dispersion is the
consequence of the ``vacuum perturbations'' on the size near the
Compton wavelength in the domain around the origin of coordinates.
However, this state is localized very well. The perturbations do
not influence the expected values of observables when the mean
moment is away from $0$.

\begin{figure}[t]
\begin{center}
\includegraphics[width=\columnwidth,clip=]{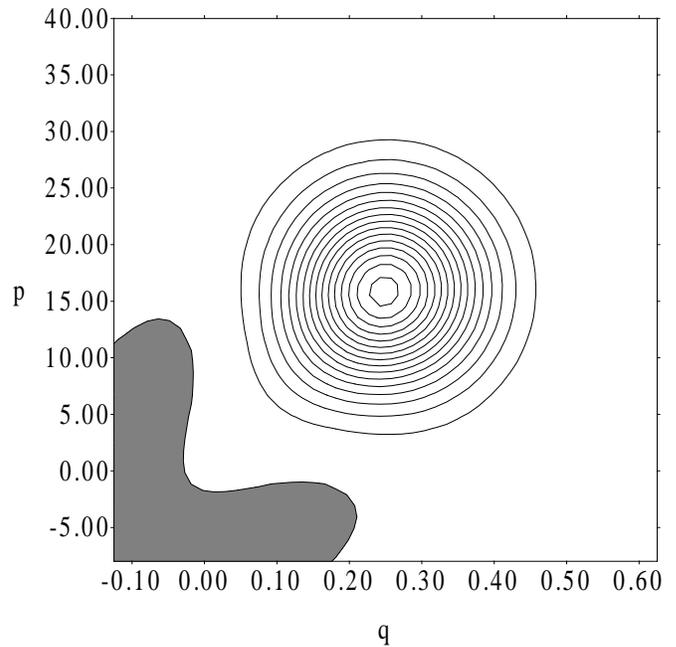}
\end{center}
\caption{\label{fig6} The contours of the Wigner function for the
coherent state of a relativistic rotator. The ratio of the Compton
wavelength and oscillator length is $\lambda=8$. The mean momentum
and coordinate are $16mc$ and $1/4\lambda_c$ respectively. The
coordinate is given in $\lambda_c$ units; the momentum is given in
$mc$ units (absolute). The domain of the negative values is marked
in black.}
\end{figure}

The Wigner function for the coherent state (\ref{f46}) of the
relativistic rotator can be written as the following expression
\cite{b18}:
\setlength\arraycolsep{2pt}
\begin{eqnarray}
&&W_{[+],\alpha}(p,q)=\frac{1}{\pi}\exp\left(-q^2-p^2\right)\exp
\left(-\left|\alpha\right|^{2}\right)\nonumber
\\ &&\times\!\sum\limits_{m,n=0}^{\infty}\!
\frac{\varepsilon(m,n)}{[\varepsilon(m)]![\varepsilon(n)]!}\left(\sqrt{2}(q+ip)
\overline{\alpha}\right)^{m}\!\left(\sqrt{2}(q-ip)\alpha\right)^{n}\nonumber\\
&&\times\!\sum\limits_{k=0}^{\min(m,n)}\!
\frac{1}{(-2(q^2+p^2))^{k}(m-k)!(n-k)!k!},\label{f70}
\end{eqnarray}
where $\varepsilon(n,m)$ is the $\varepsilon$ factor of two
variables:
\begin{equation}
\varepsilon(n,m)=\frac{{E(n)+E(m)}}{{2\sqrt
{E(n)E(m)}}}.\label{f71}
\end{equation}
The plot of this Wigner function is presented in Fig.~\ref{fig6}.
Unlike the case of a free particle, ``vacuum perturbations''
increase as $\alpha\rightarrow\infty$ here. They may be seen as
the large negative domain near the origin of coordinates.

\section{Resolution of unity}
\label{s7}

The resolution of unity is one of the key properties of the
coherent states. In this section we consider it for the
relativistic case presented here. Note that for a free particle,
where these states are the standard coherent states, the measure
is determined by Eq. (\ref{f5}). Hence we consider more
complicated cases of a relativistic rotator.

It is known (see, for example, \cite{b7}) that the integral
measure in Eq. (\ref{f4}) for the coherent states (\ref{f46}) can
be written in the form
\begin{equation}
d\mu\left(\alpha\right)=\mathcal{N}\left(\left|\alpha\right|^2\right)
\mathcal{W}\left(\left|\alpha\right|^2\right)\frac{d\alpha^{\prime}d\alpha^{\prime\prime}}
{\pi},\label{f72}
\end{equation}
where the weight function $\mathcal{W}\left(x\right)$ is
determined from the following integral equation:
\begin{equation}
\int\limits_{0}^{+\infty}x^n\mathcal{W}\left(x\right)d
x=n!\left[\varepsilon^2(n)\right]!.\label{f73}
\end{equation}

The resolution of unity for the states given in Eq. (\ref{f65}) is
rather more complicated than usual. We do not develop that
resolution of unity in this paper, but we hope to return to this
question in a subsequent article.

\section{Conclusions}
\label{s9}

In this paper we have considered the relativistic coherent state
taking into account the fact that eigenfunctions of the standard
coordinate and momentum operators have both charge components.
However, the coherent states presented here contain only one
charge component and in that case the real and imaginary parts of
the parameter $\alpha$ are uniquely related to the expected values
of the standard coordinate and momentum. We obtained this result
through the determination of the coherent states as eigenstates of
the even part of the annihilation operator. Indeed, these coherent
states do not satisfy the property of temporal stability, and time
evolution of the mean position and momentum is, generally
speaking, a nontrivial question.

In order to obtain the even part of the annihilation operator, one
needs to change the standard coordinate to the Newton--Wigner
position for the case of a free particle. Hence, the average
coordinate and momentum have no peculiarities here. Nevertheless,
the second moments (dispersion) differ from ones in the nonlocal
theory and nonrelativistic quantum mechanics, especially for very
localized states.

Both the coordinate and momentum have a nontrivial charge
structure in the case of a relativistic rotator. Furthermore, the
even parts of these operators do not satisfy the commutation
relations of the usual Heisenberg--Weyl algebra. ``Mean
positions'' have the property of a deformed algebra in this case.
Therefore, the coherent states have some peculiarities and are the
so-called nonlinear coherent states. This deformation is the
consequence of the interaction with the vacuum. In fact it is very
small, and leads to insignificant differences for the mean
trajectory, but it leads to essential peculiarities for the second
moments.

A specific peculiarity for the case of a particle in a constant
homogeneous magnetic field is the fact that the ``mean positions''
of the translational and rotational motions do not commutate with
each other. Hence, we cannot construct the coherent states that
satisfy our conditions here. However, we have presented states,
which first of all, satisfy condition~(\ref{i2}) from the
Introduction and, second, satisfy condition~(\ref{i1}) only for
one degree of freedom.

Furthermore, it should be noticed that relativistic quantum motion
has other peculiarities, which are not consequences of the
nontrivial charge structure of the coordinate and momentum
operators. These properties result from the Ehrenfest theorem
because the relativistic Hamiltonian is effectively nonquadratic.
For a free particle it leads to the effective increase of the mass
(effective mass can be presented here). For the rotational motion
it leads to the low-frequency fluctuations of the gyration radius.

\begin{acknowledgments}
The authors thank K.A. Penson and J.-M. Sixdeniers for fruitful
discussion of this work and for their very useful comments.
\end{acknowledgments}


\begin{thebibliography}{99}
\bibitem{b13} B.I. Lev, A.A. Semenov  and C.V. Usenko, Phys. Let.
A. {\bf 230}, 261 (1997). In this reference the mistaken statement
that states (\ref{f65}) are eigenstates of the both annihilation
operators, has been made. However, it does not influence the
results of that work because the effects, related to the
$\varepsilon$ factor, have not been considered there.
\end{thebibliography}
\end{document}